\documentclass[onecolumn,preprintnumbers,amsmath,amssymb,nofootinbib,prd]{revtex4}

\usepackage{graphicx}
\usepackage{dcolumn}
\usepackage{bm}
\usepackage{epsfig}
\usepackage{amsmath}
\usepackage{amssymb}

\newcommand{\be}{\begin{equation}}
\newcommand{\ee}{\end{equation}}
\newcommand{\bea}{\begin{eqnarray}}
\newcommand{\eea}{\end{eqnarray}}
\newcommand{\bdm}{\begin{displaymath}}
\newcommand{\edm}{\end{displaymath}}
\newcommand{\beas}{\begin{eqnarray*}}
\newcommand{\eeas}{\end{eqnarray*}}
\newcommand{\bml}{\begin{subequations}}
\newcommand{\eml}{\end{subequations}}

\newcommand{\kh}{\hat{k}}                 
\newcommand{\kv}{\mathbf{k}}              

\begin{document}

\preprint{ICG 05/20}

\title{Non-Gaussianity from Cosmic Magnetic Fields}

\author{Iain Brown}
\email{Iain.Brown@port.ac.uk}
\author{Robert Crittenden}%
\email{Robert.Crittenden@port.ac.uk}
\affiliation{%
Institute of Cosmology and Gravitation, University of Portsmouth, Portsmouth, UK
}%

\date{\today}

\begin{abstract}
Magnetic fields in the early universe could have played an important role in sourcing cosmological perturbations.  While not the dominant source, even a small contribution might be traceable through its intrinsic non-Gaussianity.  Here we calculate analytically the one, two and three point statistics of the magnetic stress energy resulting from tangled Gaussian fields, and confirm these with numerical realizations of the fields.  We find significant non-Gaussianity, and importantly predict higher order moments that will appear between the scalar, vector and tensor parts of the stress energy (e.g., scalar-tensor-tensor moments).  Such higher order cross correlations are a generic feature of non-linear theories and could prove to be an important probe of the early universe.
\end{abstract}

\maketitle

\section{Introduction}

It appears increasingly likely from cosmological data such as the cosmic microwave background (CMB) that the dominant source of perturbations in the universe was adiabatic and Gaussian in nature, consistent with the expectations of an inflationary universe \cite{WMAP-Bennett,WMAP-Komatsu}. However, this does not exclude the possibility that non-linear sources might also have played a role in sourcing perturbations.  While their impact on the power spectrum might be small, they could dominate any non-Gaussianities that we observe.  In order to search for these optimally we need clear predictions for the nature of the non-Gaussian signals that they would source.  Here we begin to investigate the signatures for a particular non-linear model, early-universe magnetic fields.

Magnetic fields are observed on many scales in the cosmos, from planetary scales with fields of strengths of a few Gauss, to galactic scales at a strength of approximately $\mu G$. They also exist between galaxies and are likely to be in clusters, both with field strengths somewhere between the nano- and micro-Gauss level and a coherence length on the order of megaparsecs. While fields on supercluster scales are extraordinarily difficult to detect, there are suggestions that fields up to the order of $\mu G$ may exist even there. (See for example \cite{Kronberg94, KimKronbergTribble91, ZweibelHeiles97, GrassoRubenstein01, Widrow02, Giovannini04-Mag} for reviews.)

The origin of these fields is uncertain.  Possible creation mechanisms can be separated into processes occurring before, during or after recombination.  Suggested post-recombination processes include the dynamo mechanism \cite{GrassoRubenstein01, ZeldovichRuzmaikinSolokoff80} (wherein magnetic energy is provided by rotational kinetic energy), or the adiabatic compression of an already-magnetized cloud \cite{GrassoRubenstein01}. However, these still rely on a seed field, possibly created by some large battery mechanism; an efficient dynamo could amplify a field of strength $10^{-30}G$ to observed levels, while an adiabatic compression requires a seed many orders of magnitude larger.  While mechanisms certainly exist to generate this field at reionization \cite{GnedinFerraraZweibel00,LangerPugetAghanim03} or at recombination itself \cite{Hogan00, BerezhianiDolgov03}, they might also have been created before recombination or even before nucleosynthesis; such early seed fields are constrained by limits from cosmic microwave background (see \cite{BarrowFerreiraSilk97,ClarksonEtAl02,BanerjeeJedamzik04}) and nucleosynthesis \cite{GrassoRubenstein01, CapriniDurrer02}.

Here we concentrate on fields from the early universe, and there is no shortage of suggestions for how such fields might have arisen.  They could have been produced directly by inflation, or might have been generated during the electroweak symmetry breaking phase \cite{TurnerWidrow88,Ratra92,GrassoRubenstein01, DimopoulosProkopecTornkvistDavis02, Giovannini04-Mag,ProkopecPuchwein04, BambaYokoyama04, BassettEtAl01,AshoorionMann05}. There are also recent studies into generic phase transitions generating large-scale fields (e.g., \cite{BoyanovskyVegaSimionato03,BoyanovskyVega05}). Cosmic defects might also be responsible \cite{Dimopoulos98,DaviesDimopoulos05}. More recently, attention has been given to the possibility that the fields might have been created continuously in the period between lepton decoupling and recombination, through the vorticity naturally occurring at higher order in perturbation theory \cite{BetschartDunsbyMarklund04, MatarreseEtAl04, GopalSethi04, TakahashiEtAl05}.
Even after they are produced, they could evolve in the very early universe, perhaps as a result of hydromagnetic turbulence (e.g., \cite{BrandenburgEnqvistOleson96}). (Dolgov \cite{Dolgov03} provides a brief overview of many of these creation mechanisms.)

Primordial magnetic fields can have a significant impact on the cosmic microwave background.  While early treatments focussed on this effect for a Bianchi universe \cite{Jacobs69, MilaneschiFabbri85}, more modern treatments \cite{ScannapiecoFerreira97, SubramanianBarrow98, DurrerKahniashviliYates98, KohLee00, DurrerFerreiraKahniashvili01, SubramanianBarrow02, SeshadriSubramanian00, MackKahniashviliKosowsky02, CapriniDurrerKahniashvili03, SubramanianSeshadriBarrow03, BereraEtAl03, Lewis04, Giovannini04-CMB, YamazakiIchikiKajino04} consider either small perturbations around a large-scale homogeneous field or a `tangled' field configuration.  Both of these scenarios source CMB perturbations, either directly through the scalar, vector and tensor stresses, or indirectly by the density and velocity perturbations they induce in the charged proton-electron fluids.  Code for calculating the vector and tensor anisotropies generated by primordial magnetic fields was recently added to the publicly-available CAMB \cite{CAMB,Lewis04} and that for scalars has been modelled independently \cite{KohLee00, Giovannini04-CMB, YamazakiIchikiKajino04}. In addition to creating CMB anisotropies, magnetic fields can affect also CMB polarization by inducing Faraday rotation \cite{KosowskyLoeb96, ScoccolaHarariMollerach04, CampanelliEtAl04, KosowskyEtAl04}.

It would be useful to have a firm grasp on the statistical impact magnetic fields might have upon the microwave background. 
Our ultimate goal is to translate the non-Gaussianity in the magnetic stresses into a unique non-Gaussian signature in the microwave sky \cite{BrownCrittenden2}.  While there are many potential sources of CMB non-Gaussianity, including the non-linear evolution of the perturbations and gravitational lensing (for example, \cite{BartoloMatarreseRiotto04,LesgourguesEtAl04,BartoloMatarreseRiotto05}), there are also many ways a map may be non-Gaussian.  The hope is that the different physical mechanisms will each have a characteristic non-Gaussian signature which can be searched for in the CMB sky. It is known that magnetic fields can introduce non-Gaussianity in many ways,  including via Alfv\'en turbulence or secondary perturbations, and each of these can be searched for in the CMB \cite{NaselskyEtAl04,ChenEtAl04,GopalSethi05}.  In addition, helical magnetic fields can produce parity-odd correlations in the CMB polarization field \cite{KahniashviliRatra05}.  

Here we focus on the non-Gaussianity of the magnetic sources themselves, with the three point moment as the starting point for our investigations.
To this end we model the statistics of Gaussian-random magnetic fields, concentrating on the two- and three-point statistics of the magnetic stresses and the cross-correlations between the scalar, vector and tensor components, usually considered independently.  We do so by both generating realizations of Gaussian-random magnetic fields and through numerically integrating pure analytical results, most of which are presented here for the first time. Significant non-Gaussianities and cross-correlations are expected due to the quadratic nature of the magnetic stress tensor; this suspicion is confirmed and the bispectra are presented.

In section II we review the basic model of tangled primordial magnetic fields, their impact on the CMB and their statistics, detail our realizations of these fields and construct the stress-energy tensor. In section III we briefly consider the one-point statistics of the scalar pressures of the magnetic fields, evaluating the probability distribution function, the skewness and the kurtosis for both the isotropic and the anisotropic pressures. In section IV we review the two-point results (previously demonstrated in part \cite{MackKahniashviliKosowsky02}) and present the power spectra of the separate components of the stress-energy tensor, displaying the excellent agreement between the theory and the simulations. In section V we move onto the three-point moments, deriving and presenting analytical results for their bispectra and good agreement with the simulated fields. We discuss the implications of our results in section VI.

\section{Tangled magnetic fields}
Rather than considering a large-scale homogeneous magnetic field with a small inhomogeneous perturbation, we instead consider entirely inhomogeneous fields tangled on some length scale (see for example Mack \emph{et. al.} \cite{MackKahniashviliKosowsky02}). For simplicity (and to compare with the previous literature), we assume that the magnetic fields are random variables obeying a Gaussian probability distribution function; however, our realizations may be formulated more generally than this.
Motivated by linear perturbation theory, we also take the fields to be effectively frozen, with their overall energy density decreasing as radiation ($\mathbf{B}\propto a^{-2}$, where $a$ is the scale factor.) Due to the extremely high conductivity of the early universe \cite{GrassoRubenstein01,TurnerWidrow88}, we also assume that the electric components of the electromagnetic field vanish.

Magnetic fields, being a non-linear source with a full anisotropic stress, will naturally source scalar, vector and tensor perturbations. For scalar perturbations one does not expect a significant effect on large scales. However, the contribution to the temperature auto-correlation on the microwave sky can begin to dominate at an $l$ of about $1,000$ (see for example \cite{SubramanianBarrow02, Lewis04, YamazakiIchikiKajino04}). This effect comes both from the impact of the magnetic energy and anisotropic pressure directly onto the spacetime geometry and from the Lorentz forces imparted onto the coupled proton-electron fluid. One also expects a significant impact on the vector perturbations as compared to the standard picture since the magnetic fields will both directly generate vector perturbations in the spacetime and will also contribute a Lorentz force to the solenoidal component of the velocity. Prior to neutrino decoupling a primordial magnetic field acts as a source for gravitational waves; following neutrino decoupling there will also be a contribution from the neutrinos, which Lewis has shown serves to cancel much of the magnetic stress \cite{Lewis04}. Since the Lorentz forces and the stress-energy tensor are both quadratic in the magnetic field, we also expect a level of non-Gaussianity to be imprinted onto the fluid and perturbations, even for a magnetic field that is itself Gaussian.

The features of the magnetic field -- the Lorentz forces and the direct sourcing of geometric fluctuations -- are all contained within the stress-energy tensor.  We can then consider the statistics of the stress-energy tensor alone and be hopeful of characterizing the majority of the non-Gaussian effects that might impact on the CMB. The non-Gaussianity predicted from our analysis and our simulations can be projected onto the CMB sky by folding them with the transfer functions generated by the modified CAMB code \cite{CAMB, Lewis04}; while this does not include support for scalar perturbations it will give good predictions
on all scales for the $B$-mode polarization correlations which is sourced purely by vector and tensor perturbations. However, care would have to be taken to detangle these from any $B$-modes caused by the gravitational lensing of the dominant $E$-mode polarization.
\footnote{As earlier noted, magnetic fields would also cause Faraday rotation from $E$ modes into $B$ modes \cite{KosowskyLoeb96, ScoccolaHarariMollerach04, CampanelliEtAl04, KosowskyEtAl04}; support for scalar modes in a uniform magnetic field was added to CMBFast by \cite{ScoccolaHarariMollerach04} and also to CMBFast for tangled fields by \cite{KosowskyEtAl04}, both in the case of small rotations.}

\subsection{Statistics of the fields}
 We begin by specifying the statistics of the tangled magnetic fields. In Fourier space, the divergence free condition for magnetic fields implies that
\be
\label{MagneticSpectrum}
  \left<\mathbf{B}_a(\kv)\mathbf{B}_b^*(\mathbf{k}')\right>
  =\mathcal{P}(k)P_{ab}(\mathbf{k})\delta(\kv-\mathbf{k}')+\frac{i}{2}\mathcal{H}(k)\epsilon_{abc}\hat{k}_c
\ee
where $\mathcal{P}({k})$ is the magnetic field power spectrum, $P_{ab}$ is the operator projecting vectors and tensors onto a plane orthogonal to $\kh_a$ and $\kh_b$,
\be
\label{ProjectionOperator}
  P_{ab}(\mathbf{k})=\delta_{ab}-\kh_a\kh_b ,
\ee
with $\delta_{ab}=\mathrm{diag}(1,1,1)$ and $\mathcal{H}(k)$ is the power spectrum of the anti-symmetric helical term (see for example \cite{PogosianEtAl01,CapriniDurrerKahniashvili03,KahniashviliRatra05}.) Here we have assumed the fields are statistically isotropic and homogeneous. If the magnetic fields are Gaussianly distributed, then all their statistics are determined by their power spectrum. In the interests of simplicity we henceforth assume that the helical component of the field vanishes.

In real space, the magnetic correlation function is the transform of the magnetic power spectrum, implying that
\be
\left<\mathbf{B}_a(0)\mathbf{B}_b(\mathbf{x})\right>=
\delta_{ab}C_0(x)+\frac{\partial}{\partial x_a}\frac{\partial}{\partial x_b}C_1(x)
\ee
where
$C_0(x)=\frac{V}{(2\pi)^3}\int d^3{\bf k}\mathcal{P}(k)e^{-i{\bf k\cdot x}}$ and $C_1(x)=\frac{V}{(2\pi)^3}\int d^3 {\bf k}(\mathcal{P}(k)/k^2)e^{-i{\bf k\cdot x}}$. In the limit of very small separations, the correlation becomes
\be
\left<\mathbf{B}_a(0)\mathbf{B}_b(\mathbf{x})\right>=\frac{2}{3}\delta_{ab}C_0(0)+x_ax_bC_2(0)
\ee
where $C_2(0) \propto \int dk\, k^4 \mathcal{P}(k)$ is another correlation function related to the power spectrum.

Often in the literature, the power spectrum is taken to be a simple power law,
\be
  \mathcal{P}({k})=Ak^n.
\ee
To avoid divergences, these power law spectra are generally assumed to have some small scale cutoff associated with the photon viscosity damping scale; see Mack \emph{et. al.} \cite{MackKahniashviliKosowsky02} for an evaluation of these for vector and tensor perturbations generated by a stochastic field with a power-law spectrum.  Durrer and Caprini \cite{CapriniDurrer02, DurrerCaprini03} demonstrate that, for a causally-generated magnetic field, the spectral index must be $n\geq 2$; it must also in any event be $n\geq -3$ to avoid over-production of long-range coherent fields. They also show that, for the tensor case at least, the anisotropic stress will have a white-noise ($n_{\mathrm{eff}}=0$) spectrum for all $n\geq-3/2$. They also demonstrate that nucleosynthesis limits on the gravitational waves produced by the magnetic fields place extremely strong bounds on magnetic fields, to the level of $10^{-39}G$ for inflation-produced fields with $n=0$, although this has been contested (\cite{KosowskyEtAl04, CapriniDurrer05}). For spectra that might realistically imprint on the microwave background we must consider spectra with $n<-2$, which are much less tightly constrained. For purposes of comparison we shall usually take either $n=0$ for a flat spectrum -- where ``flat'' refers to the spectrum of primordial magnetic fields themselves -- in which each mode contributes equally, or $n=-2.5$ for a spectrum nearing the ``realistically observable'' $n\approx-3$, which we shall refer to as ``steep''. Such fields can be produced by inflation (for example, \cite{Ratra92,BambaYokoyama04}).

It is worth emphasising that, while we are restricting ourselves to a power-law spectrum, this is not a necessity. We are generating simulated fields and calculating the statistics from these; it is as simple to employ other power spectra as it is to employ a power law. It is also a simple matter to employ a non-sharp damping scale -- we could, for example, employ an exponentially or Gaussian-damped tail above a particular scale $k_D$ and gain some greater freedom in modelling the effective microphysics. As a particular example, Matarrese \emph{et. al.} \cite{MatarreseEtAl04} derived the power spectrum they expect from a field sourced from second-order vorticity in the electron-baryon plasma. This power spectrum has no simple functional form but is instead presented as a numerical integration and, in principle, it would be simple to employ this integration as the input power spectrum.

The normalization of the power spectrum is typically fixed by reference to a particular co-moving smoothing scale $\lambda$ and the variance of the field strength at this scale, $B_\lambda$. Specifically, we smooth the field by convolving it with the Gaussian filter
\be
  f(k)=\mathrm{exp}\left(-\frac{\lambda^2k^2}{2}\right)
\ee
and define the variance of the field strength at the scale $\lambda$ by
\be
  B_\lambda^2=\left<B_a(\mathbf{x})B_a(\mathbf{x})\right>,
\ee
implying that the power spectrum and $B_\lambda$ are related by
\be
  B_\lambda^2=\int d^3\mathbf{k}\mathcal{P}({k})e^{-\lambda^2k^2}.
\ee
This allows us to relate the astronomically observed field strengths at, say, cluster scales, to the amplitude of the magnetic power spectrum.

For simplicity and concreteness, we will assume throughout that the fields themselves are Gaussian (consistent with previous literature). Obviously, any conclusions about the non-Gaussian signatures of magnetic fields will depend sensitively on this assumption, and we plan to explore more general scenarios in future work. Another ansatz for the fields is that they possess a $\chi^2$ probability distribution function; such fields might arise from weakly non-linear effects since they are sampled from the combination of two (first-order) Gaussian variables.  For example, the sourcing of second-order vorticity in the electron-baryon plasma by first-order density perturbations could lead to a weak but continually generated magnetic field \cite{BetschartDunsbyMarklund04, MatarreseEtAl04, GopalSethi04, TakahashiEtAl05}. Magnetic fields sourced at recombination and later are unlikely to be reasonably described by the above formalism; those sourced at recombination \cite{BerezhianiDolgov03} for example naturally arise at a very small scale.  The question of the statistical nature of both recombination and very early universe fields has not been well explored in the literature. Here we present results from a Gaussian field as an illustrative example, but further work is needed to explore the consequences of any given microphysical mechanism.

\subsection{Realizations of magnetic fields}
 To aid our study of the non-Gaussian properties of tangled magnetic fields, we create static realizations of the fields numerically.  We create the fields on a grid in Fourier space of size $l_{\mathrm{dim}}^3$, where $l_{\mathrm{dim}}$ is typically 100-200. The divergence free condition means we can generate the three magnetic field components for each $k$ mode using two complex Gaussian uncorrelated random fields with unit variance,
\bdm
 \mathbf{C}=\left(\begin{array}{c}C_1\\C_2\end{array}\right).
\edm
We then determine the magnetic field Fourier components by applying a rotation matrix,
\bdm
  \mathbf{B}=\left(\begin{array}{c}B_x\\B_y\\B_z\end{array}\right)=\mathbf{R}\cdot\mathbf{C},
\edm
where $\mathbf{R}$ is a $3 \times 2$ matrix. From the definition of the magnetic field power spectrum, we see that to get the proper statistical properties, we require
\bdm
  \left<B_{a}B_{b}\right>=R_{am}\left<C_mC_n\right>R^T_{nb}=(\mathbf{R}\cdot\mathbf{R}^T)_{ab}=\mathcal{P}(k)P_{ab}(\mathbf{k}).
\edm
\noindent While this does not specify the rotation matrix uniquely, it is straightforward to show that choosing the rotation matrix as
\be
\label{R}
 \mathbf{R}=\frac{\mathcal{P}(k)^{1/2}}{\sqrt{\kh^2_x+\kh^2_y}}\left(\begin{array}{cc}
   \kh_x\kh_z & \kh_y \\
   \kh_y\kh_z & -\kh_x \\
   -\left(\kh_x^2+\kh_y^2\right) & 0
  \end{array}\right)
\ee
will produce fields with the correct statistical properties. This rotation is well defined except in the case when $\kh_x=\kh_y=0.$  Here, $B_z=0$ and the other components are uncorrelated, so we instead choose
\be
  \label{R0} \mathbf{R}_0=\mathcal{P}(k)^{1/2}\left(\begin{array}{cc}1&0\\0&1\\0&0\end{array}\right) .
\ee
The reality of the fields is ensured by requiring $\mathbf{B}(\mathbf{-k})=\mathbf{B}(\mathbf{k})^*$.

Throughout, we are careful to avoid creating modes with frequencies higher than the Nyquist frequency of the grid, $k_{\mathrm{Nyquist}}$, which could be  into power on other frequencies. Since the quantity of greatest interest, the stress-energy, is a quadratic function of the fields, it typically will have power up to twice the cutoff frequency of the magnetic fields.  To avoid having aliasing of these fields, we generally require that the magnetic field cutoff frequency be less than half the Nyquist frequency. We also have an infra-red cut-off which is the inevitable result of working on a finite grid.

\begin{center}
\begin{figure}
\includegraphics{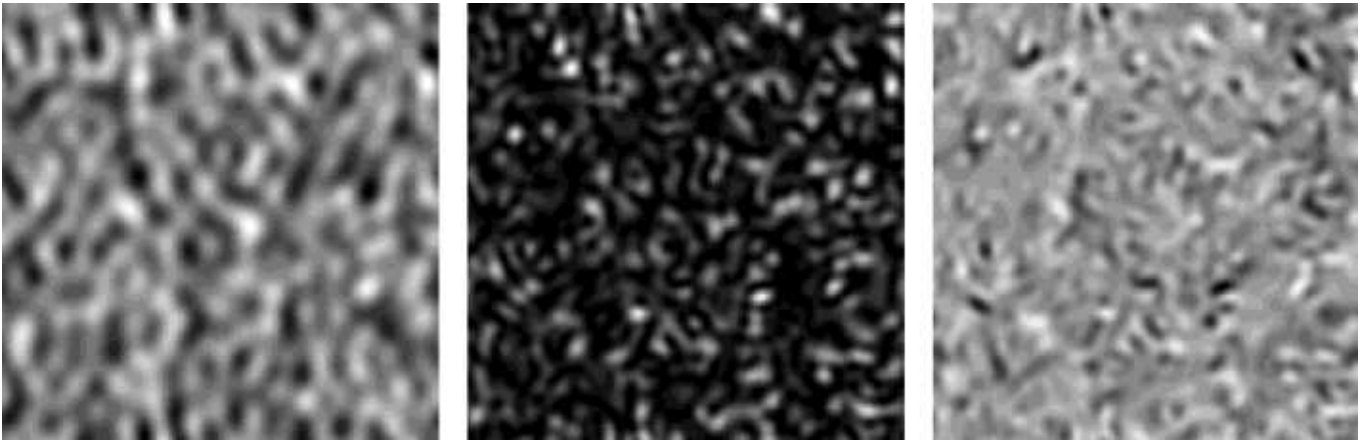}
\caption{Sample realizations of the magnetic field for a spectrum $n=0$. Left: Gaussian magnetic field slice, $B_{x}|_{z=0}$; center: the (very non-Gaussian) isotropic pressure $\tau|_{z=0}$; right: the (slightly non-Gaussian) anisotropic pressure $\tau^S|_{z=0}$.  Compared to the magnetic field, non-linearity transfers power to smaller scales in the sources. }\label{FieldSlices}
\end{figure}
\end{center}

Figure \ref{FieldSlices} shows a sample Gaussian realization of one component of the magnetic field along a slice through the realization, as well as the resulting trace and traceless components of the stress-energy (the isotropic and anisotropic pressures). Both the isotropic and anisotropic pressures show power on smaller scales than the fields themselves, a direct result of their non-linearity.  In addition, the isotropic pressure is darker, reflecting a paucity of positive fluctuations and a significant deviation from Gaussianity. The anisotropic pressure appears to be more similar to the magnetic field -- that is, relatively Gaussian. These observations will be made concrete in the next section when we consider the one-point statistics of the pressures.

\subsection{The stress-energy tensor}
 Gravitational waves are sourced directly by the stress-energy of a magnetic field, and the effect from the Lorentz force on the magnetized CMB also depends directly upon this \cite{Lewis04}; for this reason we shall concentrate our study on the space-space part of the magnetic stress-energy tensor, the stresses. In real space, these are
\be
  \tau_{ab}(\mathbf{x})=\frac{1}{2}B_i(\mathbf{x})B_i(\mathbf{x})\delta_{ab}-B_a(\mathbf{x})B_b(\mathbf{x}).
\ee
\noindent In Fourier space, the product of two magnetic field components in real space becomes a convolution,
\be
  \tilde{\tau}_{ab}(\mathbf{k})=\int B_a(\mathbf{q})B_b(\mathbf{k}-\mathbf{q})d^3\mathbf{q}.
\ee
The full stress-energy tensor in Fourier space is
\be
  \tau_{ab}(\mathbf{k})=\frac{1}{2}\delta_{ab}\tilde{\tau}_{ii}(\mathbf{k})-\tilde{\tau}_{ab}(\mathbf{k}).
\ee

Even though the stress-energy is non-linear in the fields, we will be assuming throughout that the stress-energy perturbations in the magnetic fields are small compared to the total energy density. Thus, our treatment of the perturbations induced by the magnetic fields in the photons, baryons and so forth are purely linear. In this regime, the evolution of the scalar, vector and tensor perturbations decouple and we will thus decompose the magnetic field sources into these various pieces (interpreted as the isotropic pressure, anisotropic pressure, vortical and anisotropic stresses respectively) as
\bea
  \tau_{ab}&=&\frac{1}{3}\delta_{ab}\tau+(\kh_a\kh_b-\frac{1}{3}\delta_{ab})\tau^S \nonumber \\ &&
  +2\kh_{(a}\tau_{b)}^V+\tau_{ab}^T.
\eea
\noindent We do this in Fourier space by applying combinations of projection operators;
\bea
  \tau(\mathbf{k})&=&\delta_{ab}\tau_{ab}(\mathbf{k}), \nonumber \\
\label{Projections}
  \tau^S(\mathbf{k})&=&\left(\delta_{ab}-(3/2)P_{ab}(\mathbf{k})\right) \tau_{ab}(\mathbf{k})\equiv Q_{ab}(\mathbf{k})\tau_{ab}(\mathbf{k}), \\
  \tau_a^V(\mathbf{k})&=&\kh_{(i}P_{j)a}(\mathbf{k})\tau_{ij}(\mathbf{k}) \equiv \mathcal{V}_{aij}(\mathbf{k})\tau_{ij}, \nonumber \\
  \tau_{ab}^T(\mathbf{k})&=&\left(P_{a(i}(\mathbf{k})P_{j)b}(\mathbf{k})-\frac{1}{2}P_{ij}(\mathbf{k})P_{ab}(\mathbf{k})\right)
  \tau_{ij}(\mathbf{k})\equiv \mathcal{T}_{abij}(\mathbf{k})\tau_{ij}(\mathbf{k})  \nonumber.
\eea
\noindent In the above, $(a\ldots b)$ denotes symmetrization in $a$ and $b$, \emph{i.e}, $A_{(ab)}=(1/2)(A_{ab}+A_{ba})$, and we are working with symmetrized projectors for the vector and tensor components for future ease. Here, $\kh_i\tau_i^V=\kh_i\tau_{ij}^T=\tau_{ii}^T=0.$ The first term in the stress-energy contributes purely to the scalar-trace part. For the others we can simply replace $\tau_{ab}$ with $-\tilde{\tau}_{ab}$.

The transfer functions which describe how fluctuations in the magnetic field stress energy impact the microwave background have been previously evaluated, in various semi-analytical limits in \cite{KohLee00,DurrerFerreiraKahniashvili01,MackKahniashviliKosowsky02,SubramanianBarrow02} for the temperature perturbations, for example. However, the simplest route towards folding the expected non-Gaussianities onto the CMB, will likely be from the CAMB code as modified recently by Lewis \cite{CAMB,Lewis04}; this produces the transfer functions generated by a magnetic field sourced before neutrino decoupling, neglecting the impact of the scalar perturbations. This could be supplemented by incorporating the extensions to scalar modes of Giovannini \cite{Giovannini04-CMB} into, for example, the CMBFast code \cite{SeljakZaldariagga96}. We leave this issue to a later paper.

\section{One-Point Moments}
 There are many ways to characterize non-Gaussianity, particularly given such a strongly non-linear stress-energy term. In this section we briefly consider the skewness and kurtosis of the one-point probability distributions of the isotropic and anisotropic pressures. In this section the results we present are the mean of twenty realizations with a grid-size of $l_{\mathrm{dim}}=192$, and the errors quoted are one standard deviation.

The simplest to consider is the distribution of the trace part, since it is simply the square of the magnetic field.  Despite the divergence-free condition, the three components of the magnetic field at a single point in space are uncorrelated and Gaussian (as was shown above.) Thus we expect the trace of the stress-energy to have a $\chi^2$ distribution with three degrees of freedom.

All the moments of a one-point distribution may be given by its moment generating function as
\be
  \mu'_n \equiv \langle X^n \rangle = \frac{\partial^n}{\partial t^n}\left.M(t)\right|_{t=0}
\ee
where, for a $\chi^2$ distribution with $p$ degrees of freedom
\be
  M(t)=\frac{1}{\left(1-2t\right)^{p/2}}.
\ee
The central moments are then readily calculated and the normalized skewness and kurtosis are defined to be
\be
  \gamma_1=\frac{\mu_3}{\mu_2^{3/2}}, \quad \gamma_2=\frac{\mu_4}{\mu_2^{\phantom{2}2}}-3
\ee
\noindent We quickly find that, for the $\chi^2$ distribution, the normalized skewness and kurtosis are
\be
  \gamma_1=\sqrt{\frac{8}{p}}\approx 1.633, \quad \gamma_2=\frac{12}{p}=4
\ee
\noindent where the numerical results are for a distribution with $3$ degrees of freedom.  The results from the realizations can be seen to be in agreement with the predictions; with a flat spectrum we find that, for the isotropic pressure, $\gamma_1=1.63\pm 0.01$ and $\gamma_2=3.99\pm0.05$. For a more realistically observable field, with a power spectrum of $n=-2.5$, say, we find $\gamma_1=1.61\pm0.01$ and $\gamma_2=3.92\pm0.05$. It is apparent that the statistics for the isotropic pressure are, as expected, relatively insensitive to the spectral index one employs.

The anisotropic stress is harder to characterize because it is not a local function of the fields, but contains derivatives of them.  However, it effectively is the sum of the products of two Gaussian fields which are, for the most part, independent of each other.  The distribution of the product of two independent Gaussians is non-Gaussian but is symmetric (actually following a modified Bessel distribution, as shown in the appendix of \cite{BoughnCrittenden05}.)  Thus the effect of adding such terms is to dilute the skewness.  That is, while the isotropic stress is the sum of three very skewed chi-squared distributed variables, the anisotropic stress is the sum of chi-squared terms and symmetric modified Bessels, making the result less skewed.

\begin{figure}
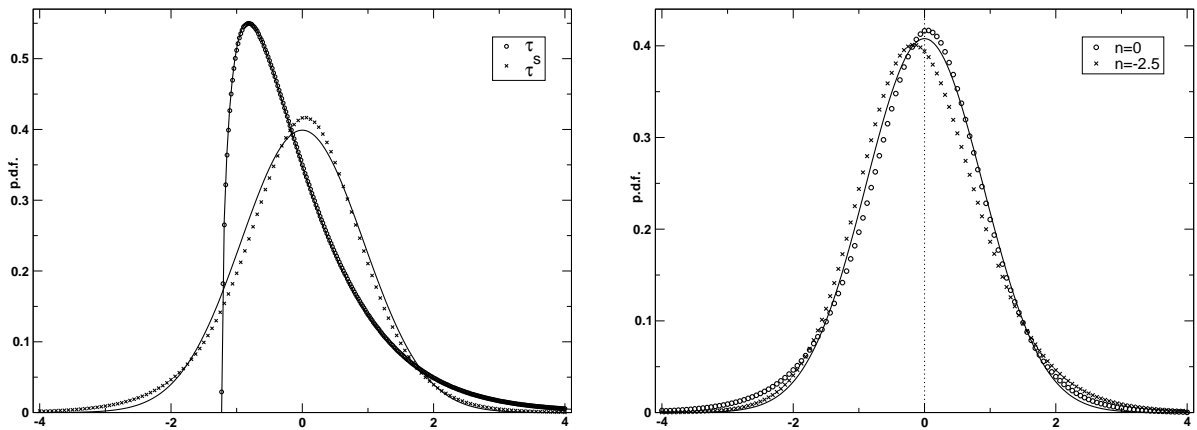
\begin{center}
\includegraphics{PDFs_n0.eps}\qquad\includegraphics{PDFs_comparison.eps}
\caption{The left figure shows the probability distribution of the isotropic and anisotropic pressures for a spectrum $n=0$, with a Gaussian distribution shown for comparison.  The isotropic distribution is well fit by a $\chi^2$ curve.
The right figure compares the anisotropic pressure distribution for different spectral indices. The $x$-axis is in units of the root mean square amplitude of the relevant field. }\label{PDFs}\end{center}\end{figure}

The probability distributions of the isotropic and anisotropic stresses for a flat spectrum are plotted in the left panel of Figure \ref{PDFs} along with a Gaussian and a $\chi^2$ distribution.  The damping scale we employed was $k_c=l_{\mathrm{dim}}/4.1$ -- \emph{i.e.} just beneath half the Nyquist frequency.  For the steep spectra, we also used twenty realizations and a grid-spacing of $l_{\mathrm{dim}}=192$ but with a damping scale at the size of the grid to ensure a reasonable mode coverage in the low-$k/k_c$ regime.
The anisotropic pressure distribution has quite different properties when the spectral index is changed, including a switch in sign of the skewness. For the flat spectrum we find $\gamma_1=-0.24\pm0.003$ and $\gamma_2=1.10\pm0.01$, while with a steep spectrum with a power spectrum of $n=-2.5$, we find  $\gamma_1=0.38\pm0.01$ and $\gamma_2=0.86\pm0.02$.
The distributions are plotted in the right-hand side of the figure, again with a sample Gaussian, and the change in the skewness is readily apparent.

\section{Two-Point moments}
 We next calculate the two point power spectra of the various perturbation types. These give a useful example of how the higher order calculations will proceed and give a means of testing our realizations. Some of these have been previously calculated, such as the vector \cite{MackKahniashviliKosowsky02} and tensor \cite{DurrerFerreiraKahniashvili01, MackKahniashviliKosowsky02, CapriniDurrer02} power spectra, while the trace and traceless scalar auto-correlations and their cross correlation, to our knowledge, have not.

By the nature of the scalar-vector-tensor decomposition, we do not expect any cross correlations except between the trace and traceless scalar pieces. Thus, we consider five power spectra: one cross spectrum and the auto-spectra of the four pieces of the stress-energy.  We focus on constructing rotationally invariant spectra which will contain all the information in the general correlations.

In general, the power spectra will involve expectations of four magnetic fields. Since these are assumed to be Gaussian, they can be evaluated using Wick's theorem which, for four Gaussian fields, may be expressed as
\bdm
  \left<ABCD\right>=\left<AB\right>\left<CD\right>+\left<AC\right>\left<BD\right>+\left<AD\right>\left<BC\right> .
\edm
It is most useful to begin with the general two point correlation,
\bea
  \lefteqn{\left<\tilde{\tau}_{ab}(\mathbf{k})\tilde{\tau}_{cd}^*(\mathbf{p})\right>=\delta(\mathbf{k}-\mathbf{p}) \int d^3\mathbf{k}'\mathcal{P}(k')\mathcal{P}(\left|\mathbf{k-k'}\right|)} \\ && \qquad
 \times \left(P_{ac}(\mathbf{k'})P_{bd}(\mathbf{k-k'})
 +P_{ad}(\mathbf{k'})P_{bc}(\mathbf{k-k'})\right). \nonumber
\eea
\noindent (Note that there are two terms rather than three since we interested in the perturbations from the mean value of the field.) The power spectra of the various stresses may be obtained from this by applying different combinations of the projection operators (\ref{Projections}) to yield,
\be
  \left<\tau_A(\mathbf{k})\tau_B(\mathbf{p})\right>=\delta(\mathbf{k}-\mathbf{p})
  \int d^3\mathbf{k}'\mathcal{P}(k')\mathcal{P}(\left|\mathbf{k-k'}\right|)
  \mathcal{F}_{AB}
\ee
\noindent with $A$ and $B$ denoting the two stress components and $\mathcal{F}_{AB}=\mathcal{F}_{AB}(\gamma,\mu,\beta)$ denoting the relevant angular integrand.  The relevant angles possible between the wavevectors have been defined as
\be
  \gamma=\hat{\mathbf{k}}\cdot\hat{\mathbf{k}}', \quad \mu=\hat{\mathbf{k}}'\cdot\widehat{\mathbf{k-k'}}, \quad \beta=\hat{\mathbf{k}}\cdot\widehat{\mathbf{k-k'}} ,
\ee
where $\hat{\mathbf{k-k'}}$ denotes the unit vector in the direction of $\mathbf{k-k'}$.

The trace-trace correlation is found by applying the operator $(1/4)\delta_{ab}\delta_{cd}$ whence we obtain
\be
  \mathcal{F}_{\tau\tau}=\frac{1}{2}\left(1+\mu^2\right) .
\ee
Similarly, we obtain the traceless scalar auto-correlation function by applying $(-1)^2Q_{ab}(\mathbf{k})Q_{cd}(\mathbf{k})$; some algebra yields the result
\be
  F_{\tau^S\tau^S}=2+\frac{1}{2}\mu^2-\frac{3}{2}\left(\gamma^2+\beta^2\right)-{3}\gamma\mu\beta+\frac{9}{2}\gamma^2\beta^2.
\ee
The cross correlation between the trace and traceless scalar pieces requires the operator $-(1/2)\delta_{ab}Q_{cd}(\mathbf{k})$. This yields
\be
  F_{\tau\tau^S}=-1+\frac{3}{2}\left(\gamma^2+\beta^2\right)+\frac{1}{2}\mu^2-\frac{3}{2}\mu\gamma\beta.
\ee

For the vector and tensor contributions, it is useful to construct rotationally invariant combinations that can be relatively easily mapped onto the CMB. The divergenceless condition on the vectors implies that their correlation function can be written as
\be
\label{SpectrumTv}
  \left<\tau^V_a(\mathbf{k})\tau^{*V}_b(\mathbf{p})\right>=\frac{1}{2}
  \mathcal{P}^V(k)P_{ab}(\mathbf{k})\delta(\mathbf{k}-\mathbf{p}) .
\ee
where our definition differs by a factor of two from Mack \emph{et. al.}. All the information is condensed in the rotationally invariant vector isotropic spectrum $\mathcal{P}^V(k)=\left<\tau_a^V(\mathbf{k})\tau_a^{*V}(\mathbf{k})\right>.$ The operator necessary to recover this is $\kh_{(a}P_{b)i}(\mathbf{k})\kh_{(c}P_{d)i}(\mathbf{k})=\kh_a\kh_{(c}P_{d)b}(\mathbf{k})+\kh_d\kh_{(a}P_{b)c}(\mathbf{k})$. The resulting angular term can then be shown to be
\be
  F_{\tau^V\tau^V}=1-2\gamma^2\beta^2+\mu\gamma\beta.
\ee

Similar arguments apply for the tensor correlations. The full tensor two-point correlation is
\be
\label{SpectrumTt}
  \left<\tau_{ab}^T(\mathbf{k})\tau_{cd}^{*T}(\mathbf{p})\right>=\frac{1}{4}\mathcal{P}^T(k)\mathcal{M}_{abcd}(\mathbf{k})\delta(\mathbf{k}-\mathbf{p}),
\ee
where $\mathcal{M}_{abcd}(\mathbf{k})=P_{ac}(\mathbf{k})P_{bd}(\mathbf{k})+P_{ad}(\mathbf{k})P_{bc}(\mathbf{k})-P_{ab}(\mathbf{k})P_{cd}(\mathbf{k})$ which, as can be readily shown, satisfies the transverse-traceless condition on the tensors, and $\delta_{ac}\delta_{bd}\mathcal{M}_{abcd}(\mathbf{k})=4$. We focus on the rotationally invariant tensor isotropic spectrum $\mathcal{P}^T(k)=\left<\tau_{ij}^{T}(\mathbf{k})\tau_{ij}^{*T}(\mathbf{k})\right>$. Using the tensor projection operators and simplifying, the relevant operator is $\left(P_{i(a}(\mathbf{k})P_{b)i}(\mathbf{k})-(1/2)P_{ij}(\mathbf{k})P_{ab}(\mathbf{k})\right)\left(P_{i(c}(\mathbf{k})P_{d)i}(\mathbf{k})-(1/2)P_{ij}(\mathbf{k})P_{cd}(\mathbf{k})\right)=P_{c(a}(\mathbf{k})P_{b)d}(\mathbf{k})
-(1/2)P_{ab}(\mathbf{k})P_{cd}(\mathbf{k})$. This leads to a simple angular term of
\be
  F_{\tau^T\tau^T}=(1+\gamma^2)(1+\beta^2).
\ee

The vector and tensor results differ from those otherwise presented (see Durrer \emph{et. al.} and Mack \emph{et. al.} \cite{DurrerFerreiraKahniashvili01, MackKahniashviliKosowsky02, CapriniDurrer02}), the vector case by $\beta^2-\gamma^2$ as a result of employing the symmetrized projection, and the tensor case by $\gamma^2-\beta^2$. It is straightforward, however, to see that if one redefines the integration wavemode as $\mathbf{k}'=\mathbf{k}''-\mathbf{k}$ one maps $\mu'\rightarrow\mu,\;\beta'\rightarrow\gamma,\;\gamma'\rightarrow\beta$. On integration, then, the product $\gamma^2\beta^2$ is invariant while $\gamma^2-\beta^2$ may be taken to vanish. Our results are thus in agreement with those previously presented.

\begin{figure}\begin{center}\includegraphics{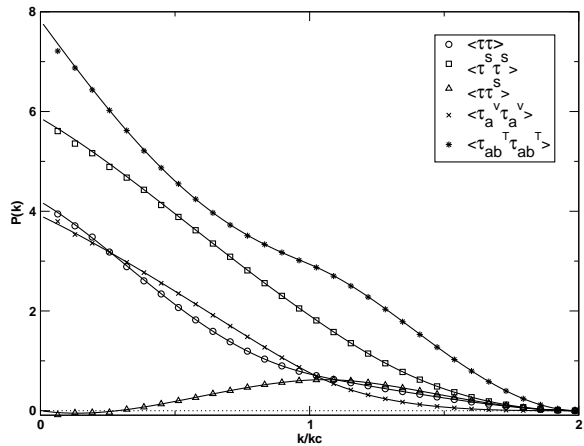}
\caption{Magnetic stress-energy power spectra ($n=0$) -- the realizations agree well with the analytic predictions. The error bars from the realisations are small except at very low $k$.}\label{Spectra_n0}\end{center}\end{figure}

We can compare numerical integrations of these power spectra with the results arising from the realised magnetic fields. Our results for a flat power spectrum ($n=0$) are presented in Figure \ref{Spectra_n0} where $P(k)$ denotes the various spectra (all presented on the same scale). The agreement between the analytic results (lines) and a simulated field (data points) is striking. We have plotted the power spectra averaged over twenty realizations with a grid-size of $l_{\mathrm{dim}}=192$, a damping scale at $k_c=l_{\mathrm{dim}}/4.1$, with error bars of 1-$\sigma$. We have also rebinned the results because otherwise the data points obscure the theory.

We could also consider magnetic fields with $n>2$ corresponding to causally-generated fields \cite{CapriniDurrer02}; the features for such fields are qualitatively similar to those for a flat spectra and there are no difficulties in evaluating the theoretical predictions in this regime. However, as commented, Caprini and Durrer demonstrated that primordial fields of this type would be unobservably small in order to not violate nucleosynthesis bounds on gravitational waves. Moreover, blue spectra naturally tend to pile power around the cut-off scale $k_c$ and since we are not at all modelling the microphysics the results would have to be treated with caution.

\begin{figure}\begin{center}\includegraphics{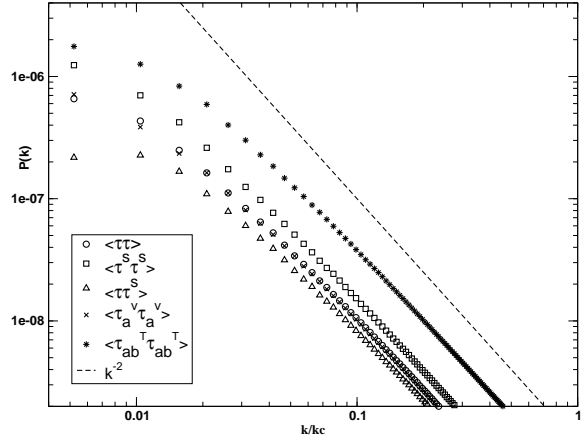}
\caption{Magnetic stress-energy power spectra ($n=-2.5$). The dashed line shows the spectral dependence expected naively, $\propto k^{2n+3}$.  The turnover at low $k$ reflects the finite size of the grid. }\label{Spectra_n-2.5}\end{center}\end{figure}

Analytic results can be found in certain limits.  There are two regimes of interest for the spectral index, as shown by Durrer et. al. \cite{DurrerFerreiraKahniashvili01}.  For $n>-3/2$ the integrations are dominated by the cutoff scale, resulting in constant spectra.  In this regime, if $k\ll k_c$, the angular integrations are straight forward ($\mu\simeq-1, \beta\simeq-\gamma$.) Relative to the trace correlation, the amplitudes of the other correlations are $\tau^S/\tau=7/5$, $\tau_\times/\tau=0$, $\tau^V/\tau=14/15$ and $\tau^T/\tau=28/15$ respectively. For $n<-3/2$ the situation is considerably more complex and we content ourselves with the results of our simulations in Figure \ref{Spectra_n-2.5}. Immediately apparent is that the effects of the cutoff are reduced by the strongly tilted magnetic spectrum, with each spectrum quickly approaching the power law, $P_A(k)\propto k^{2n+3}$, that is naively expected from the $k$-integration. Also notable is the change of behaviour of the scalar cross-correlation; whereas this vanishes on large scales in the $n>-3/2$ regime, it remains finite on large scales for $n<-3/2$ and so in principle might be observable on the sky. There are also the effects of the infrared cut-off causing a suppression at low-$k$.

\section{Three-Point Moments}
 In this section we focus on the three point moments in Fourier space, for which it is possible (if laborious) to obtain analytic expressions.  The magnetic field realizations provide a way of exploring other kinds of non-Gaussianities which may arise.

At the three point level, it is no longer guaranteed that correlations between the scalar, vector and tensor pieces will vanish, and we present some of the first calculations of these here. There are many possible three point moments, but here we consider only the rotationally invariant combinations $\langle\tau\tau\tau\rangle$, $\langle\tau\tau\tau^S\rangle$, $\langle\tau\tau^{S}\tau^{S}\rangle$, $\langle\tau^{S}\tau^{S}\tau^{S}\rangle$, $\langle\tau\tau^V_a\tau_a^V\rangle$ and $\langle\tau^S\tau^V_a\tau_a^V\rangle$. Due to their complexity we postpone calculations of the correlations with the tensor components to a later paper and content ourselves with presenting the results from the realizations; for completeness these are $\langle\tau\tau_{ab}^T\tau_{ab}^T\rangle$ and $\langle\tau^S\tau_{ab}^T\tau_{ab}^T\rangle$ for correlations with the scalars, $\langle\tau_a^V\tau_{ab}^T\tau_b^V\rangle$ with the vectors and the auto-correlation $\langle\tau_{ab}^T\tau_{bc}^T\tau_{ac}^T\rangle$. We work throughout in Fourier space, where the three-point moments are known as the bispectra. One advantage of working in Fourier space is that the transfer functions, which fold in the fluid dynamics and describe the impact on the microwave background, are local.

\subsection{General considerations}
\begin{figure}\begin{center}
\includegraphics{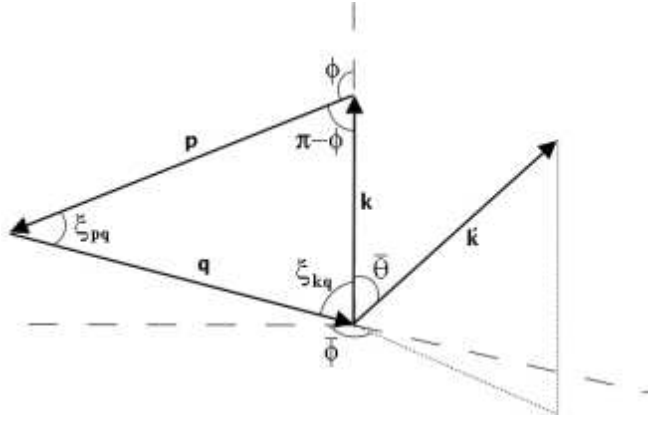}
\caption{The geometry for the bispectra calculations; $\mathbf{k},\mathbf{p},\mathbf{q}$ are the wavevectors for the three legs, while $\mathbf{k'}$ is an integration mode.}
\label{BispectraGeometry}\end{center}\end{figure}

 In principle we can calculate all the three point statistics described above in Fourier space.  The bispectra involve three wave modes, and since we assume the fields are homogeneous and isotropic, the sum of the three modes must be zero. Thus the bispectra are a function of the amplitudes of the modes alone (or alternatively, two amplitudes and the angle between them.)  We denote different geometries by selecting a baseline $\mathbf{k}$ and a vector $\mathbf{p}$ making an angle $\phi$ with $\mathbf{k}$ and having an amplitude $p=rk$ (see Figure \ref{BispectraGeometry}). We may then calculate $\mathbf{q}=-\mathbf{k}-\mathbf{p}$. For simplicity, we here concentrate on the colinear (degenerate) case in which $\mathbf{p}=\mathbf{k}$ implying $\mathbf{q}=-2\mathbf{k}$ -- that is, $r=1$ and $\phi=0$.

We calculate the bispectra analogously to the power spectra, although matters are complicated by the need to deal with expectations of six fields rather than four. The object of most general interest is $\mathcal{B}_{ijklmn}(\mathbf{k},\mathbf{p},\mathbf{q}) \equiv\left<\tilde{\tau}_{ij}(\mathbf{k})\tilde{\tau}_{kl}(\mathbf{q})\tilde{\tau}_{mn}(\mathbf{p})\right>,$ which is related to the expectation value of six magnetic fields,
\be
  \mathcal{B}_{ijklmn}(\mathbf{k},\mathbf{p},\mathbf{q})=\iiint d^3\mathbf{k}'d^3\mathbf{p}'d^3\mathbf{q}'\left<B_i(\mathbf{k}')B_j(\mathbf{k}-\mathbf{k}')B_k(\mathbf{p}')B_l(\mathbf{p}-\mathbf{p}')B_m(\mathbf{q}')B_n(\mathbf{q}-\mathbf{q}')\right> .
\ee
As in the two-point case, all three-point moments of interest may be found by applying the relevant projection operator, $\mathcal{A}_{ijklmn}$, to this. Expanding this six-point correlation with Wick's theorem generates fifteen terms, eight of which contribute to the reduced bispectrum, that is, the bispectrum neglecting the one-point terms proportional to $\delta(\mathbf{k})$, $\delta(\mathbf{p})$ or $\delta(\mathbf{q})$. This leads eventually to
\bea
\label{SixFieldCorrelation}
  \lefteqn{\mathcal{B}_{ijklmn}(\mathbf{k},\mathbf{p},\mathbf{q})=} \\ &&
  \delta(\mathbf{k}+\mathbf{p}+\mathbf{q})\int d^3\mathbf{k}'\mathcal{P}(k')\mathcal{P}(\left|\mathbf{k}-\mathbf{k}'\right|)
  \mathcal{P}(\left|\mathbf{p}+\mathbf{k}'\right|) P_{ik}(\mathbf{k}')\left(P_{jm}(\mathbf{k}-\mathbf{k}')
  P_{ln}(\mathbf{p}+\mathbf{k}')\right)
  +\left\{i\leftrightarrow j, p\rightarrow q\right\}, \{k\leftrightarrow l, m\leftrightarrow n.\} \nonumber
\eea
These eight terms reduce to the same contribution if the projection tensor $\mathcal{A}_{ijklmn}$ that recovers a set bispectrum is independently symmetric in $\{ij\}$, $\{kl\}$ and $\{mn\}$.

In the power spectra calculations, the geometry was straight forward; here it is considerably more complicated. The three wavevectors of the bispectrum are constrained by homogeneity to obey $\mathbf{k+q+p}=0$. Combined with the dummy integration wavevector, these define a four sided tetrahedron.  This has six edges, $\mathbf{k,q,p,k',k-k'}$ and $\mathbf{p+k'}$.  From these, we can generate fifteen unique angles which could come into the bispectra calculation.  This is to be compared to just three edges and three angles required for the power spectra.  Clearly these angles are not all independent; they are, in fact, functions of just five underlying angles.

For our purposes, it is easiest to work with the fifteen which we separate into four hierarchies; those between the set wavevectors $\mathbf{k}$, $\mathbf{p}$ and $\mathbf{q}$, angle cosines of these vectors with $\mathbf{k}'$, cosines with $\mathbf{k}-\mathbf{k}'$, and cosines with $\mathbf{p}+\mathbf{k}'$. The final group are defined below. We take the angles, with $\left\{\mathbf{a},\mathbf{b}\right\}\subset\left\{\mathbf{k},\mathbf{p},\mathbf{q}\right\}$, to be
\bea
  &\theta_{ab}=\hat{\mathbf{a}}\cdot\hat{\mathbf{b}}, \quad \alpha_a=\hat{\mathbf{a}}\cdot\hat{\mathbf{k}}', \quad \beta_a=\hat{\mathbf{a}}\cdot\widehat{\mathbf{k}-\mathbf{k}'}, \quad \gamma_a=\hat{\mathbf{a}}\cdot\widehat{\mathbf{p}+\mathbf{k}'},&
 \nonumber \\
  &\bar{\beta}=\mathbf{\hat{k}'}\cdot\mathbf{\widehat{k-k'}} \quad \bar{\gamma} = \mathbf{\hat{k}'}\cdot\mathbf{\widehat{p+k'}}\quad \bar{\mu} = \mathbf{\widehat{k - k'}}\cdot\mathbf{\widehat{p+k'}}.&
\eea
In terms of the angles $\xi_{kq}$ and $\xi_{pq}$ in Figure \ref{BispectraGeometry} we obviously have $\theta_{kq}=-\cos\xi_{kq}$ and similarly for $\theta_{pq}$. We also have $\alpha_k=\cos\bar{\theta}$.

In a manner entirely analogous to the two-point case we find the different bispectra by applying to (\ref{SixFieldCorrelation}) different projection operators to extract the relevant scalar, vector or tensor parts and, given that we ensure that $\mathcal{A}_{ijklmn}$ has the required symmetries, we may express the bispectra as
\be
\label{GenericBispectra}
  \left<\tau_A(\mathbf{k})\tau_B(\mathbf{q})\tau_C(\mathbf{p})\right>
    =\delta(\mathbf{k}+\mathbf{p}+\mathbf{q})\int d^3\mathbf{k}'\mathcal{P}(k')
    \mathcal{P}(\left|\mathbf{k}-\mathbf{k}'\right|)\mathcal{P}(\left|\mathbf{p}+\mathbf{k}'\right|)
    \left(8\mathcal{F}_{ABC}\right)
\ee
\noindent where $\left\{ABC\right\}$ denote denote different parts of the stress-energy tensor and $\mathcal{F}_{ABC}$ is the relevant angular component.

\subsection{Scalar bispectra}
 We begin with the simplest case, the bispectrum of the magnetic pressure. This is found by defining $\mathcal{A}_{ijklmn}=(1/8)\delta_{ij}\delta_{kl}\delta_{mn}$ which gives us
\be
  8\mathcal{F}_{\tau\tau\tau}=\bar{\beta}^2+\bar{\gamma}^2+\bar{\mu}^2-\bar{\beta}\bar{\gamma}\bar{\mu} .
\ee

The first scalar cross-correlation will be between the square of the pressure and the anisotropic pressure, found by using $\mathcal{A}_{ijklmn}=(-1/4)\delta_{ij}\delta_{kl}Q_{mn}(\mathbf{q})$ to give
\be
  -8\mathcal{F}_{\tau\tau\tau^S}=3\left(1-\alpha_q^2-\gamma_q^2-\beta_q^2-\frac{1}{3}(\bar{\beta}^2+\bar{\gamma}^2+\bar{\mu}^2)
   +\alpha_q(\beta_q\bar{\beta}+\gamma_q\bar{\gamma})+\bar{\mu}(\beta_q\gamma_q+\frac{1}{3}\bar{\beta}\bar{\gamma})
   -\bar{\beta}\bar{\gamma}\beta_q\gamma_q\right)
\ee

Similarly the second scalar cross-correlation, with $\mathcal{A}_{ijklmn}=(1/2)\delta_{ij}Q_{kl}(\mathbf{p})Q_{mn}(\mathbf{q})$, gives
\be
  \mathcal{F}_{\tau\tau^S\tau^S}=\sum_{n=0}^5\mathcal{F}_{\tau\tau^S\tau^S}^{n}
\ee
where
\bea
  8\mathcal{F}_{\tau\tau^S\tau^S}^0&=&-6, \nonumber \\
  8\mathcal{F}_{\tau\tau^S\tau^S}^1&=&0, \nonumber \\
  8\mathcal{F}_{\tau\tau^S\tau^S}^2&=&\bar{\beta}^2+\bar{\gamma}^2+\bar{\mu}^2+3\left(\alpha_p^2+\alpha_q^2+\beta_p^2+\beta_q^2
   +\gamma_p^2+\gamma_q^2\right)+9\theta_{pq}^2, \\
  8\mathcal{F}_{\tau\tau^S\tau^S}^3&=&-\left(\bar{\beta}\bar{\gamma}\bar{\mu}+3\bar{\mu}(\beta_p\gamma_p+\beta_q\gamma_q)
   +\bar{\gamma}(\alpha_p\gamma_p+\alpha_q\gamma_q)+\bar{\beta}(\alpha_p\beta_p+\alpha_q\beta_q)+9\theta_{pq}(\alpha_p\alpha_q
   +\beta_p\beta_q+\gamma_p\gamma_q)\right) \nonumber \\
  8\mathcal{F}_{\tau\tau^S\tau^S}^4&=&3\left(\bar{\beta}(\bar{\mu}\alpha_p\gamma_p+\bar{\gamma}\beta_q\gamma_q+3\alpha_p\beta_q\theta_{pq})
   +3\left(\alpha_p\gamma_p\alpha_q\gamma_q+\beta_p\gamma_p\beta_q\gamma_q\right)\right) \nonumber \\
  8\mathcal{F}_{\tau\tau^S\tau^S}^5&=&-9\bar{\beta}\alpha_p\gamma_p\beta_q\gamma_q . \nonumber
\eea

Finally the anisotropic scalar bispectrum is found by applying $A_{ijkmln}=(-1)^3Q_{ij}(\mathbf{k})Q_{kl}(\mathbf{p})Q_{mn}(\mathbf{q})$ which results in
\be
  \mathcal{F}_{\tau^S\tau^S\tau^S}=\sum_{n=0}^6\mathcal{F}_{\tau^S\tau^S\tau^S}^{n}
\ee
\noindent with
\bea
  -8\mathcal{F}_{\tau^S\tau^S\tau^S}^0&=&9 \nonumber \\
  -8\mathcal{F}_{\tau^S\tau^S\tau^S}^1&=&0 \nonumber \\
  -8\mathcal{F}_{\tau^S\tau^S\tau^S}^2&=&-\left(
   \bar{\beta}^2+\bar{\gamma}^2+\bar{\mu}^2
   +3(\alpha_k^2+\alpha_p^2+\alpha_q^2+\beta_k^2+\beta_p^2+\beta_q^2+\gamma_k^2+\gamma_p^2+\gamma_q^2)
   +9(\theta_{kp}^2+\theta_{kq}^2+\theta_{pq}^2)
   \right) \nonumber \\
  -8\mathcal{F}_{\tau^S\tau^S\tau^S}^3&=&3\left(
   \bar{\mu}(\beta_k\gamma_k+\beta_p\gamma_p+\beta_q\gamma_q+\frac{1}{3}\bar{\beta}\bar{\gamma})
   +\bar{\gamma}(\alpha_k\gamma_k+\alpha_p\gamma_p+\alpha_q\gamma_q)
   +\bar{\beta}(\alpha_k\beta_k+\alpha_p\beta_p+\alpha_q\beta_q)
    \right. \nonumber \\ && \;
   +3\theta_{kp}(\alpha_k\alpha_p+\beta_k\beta_p+\gamma_k\gamma_p)
   +3\theta_{kq}(\alpha_k\alpha_q+\beta_k\beta_q+\gamma_k\gamma_q)
   +3\theta_{pq}(\alpha_p\alpha_q+\beta_p\beta_q+\gamma_p\gamma_q)
   +9\theta_{kp}\theta_{kq}\theta_{pq}
   \bigg) \nonumber \\
  -8\mathcal{F}_{\tau^S\tau^S\tau^S}^4&=&-3\bigg(
   \bar{\gamma}\bar{\mu}\alpha_k\beta_k+\bar{\beta}\bar{\mu}\alpha_p\gamma_p+\bar{\beta}\bar{\gamma}\beta_q\gamma_q
   +3(\alpha_k\beta_k(\alpha_p\beta_p+\alpha_q\beta_q)
   +\alpha_p\gamma_p(\alpha_k\gamma_k+\alpha_q\gamma_q)
   +\beta_q\gamma_q(\beta_k\gamma_k+\beta_p\gamma_p))
    \nonumber \\ && \;
   +3(\bar{\mu}\theta_{kp}\beta_k\gamma_p+\bar{\gamma}\theta_{kq}\alpha_k\gamma_q+\bar{\beta}\theta_{pq}\alpha_p\beta_q)
   +9(\theta_{kp}\theta_{kq}\gamma_p\gamma_q+\theta_{kp}\theta_{pq}\beta_k\beta_q+\theta_{kq}\theta_{pq}\alpha_k\alpha_p)
   \bigg) \nonumber \\
  -8\mathcal{F}_{\tau^S\tau^S\tau^S}^5&=&9\left(
   \bar{\mu}\alpha_k\beta_k\alpha_p\gamma_p
   +\bar{\gamma}\alpha_k\beta_k\beta_q\gamma_q+\bar{\beta}\alpha_p\gamma_p\beta_q\gamma_q
   +3(\theta_{kp}\beta_k\gamma_p\beta_q\gamma_q+\theta_{kq}\alpha_k\alpha_p\gamma_p\gamma_q+\theta_{pq}\alpha_k\beta_k\alpha_p\beta_q)
   \right) \nonumber \\
  -8\mathcal{F}_{\tau^S\tau^S\tau^S}^6&=&-27\alpha_k\beta_k\alpha_p\gamma_p\beta_q\gamma_q. \nonumber
\eea

\subsection{Cross bispectra}
For the vector and tensor correlations we restrict ourselves to the various rotationally-invariant quantities, which can be identified with cross-correlations between the scalar pressures and either the vector or tensor moduli. The first of these, the correlation between the scalar pressure and the vorticity, we recover with the operator $\mathcal{A}_{ijklmn}=(1/2)\delta_{ij}\hat{p}_{(k}P_{l)a}(\mathbf{p})\hat{q}_{(m}P_{n)a}(\mathbf{q})$. The eventual result is
\be
  \mathcal{F}_{\tau\tau^V\tau^V}=\sum_{n=1}^6\mathcal{F}_{\tau\tau^V\tau^V}^{n}
\ee
\noindent with
\bea
  8\mathcal{F}_{\tau\tau^V\tau^V}^1&=&-3\theta_{pq} \nonumber \\
  8\mathcal{F}_{\tau\tau^V\tau^V}^2&=&\gamma_p\gamma_q \nonumber \\
  8\mathcal{F}_{\tau\tau^V\tau^V}^3&=&\bar{\mu}(\beta_p\gamma_q+\beta_q\gamma_p)+\bar{\gamma}(\alpha_p\gamma_q+\alpha_q\gamma_p)
   +2\theta_{pq}(\alpha_p^2+\beta_p^2+\gamma_p^2+\alpha_q^2+\beta_q^2+\gamma_q^2+\frac{1}{2}\bar{\beta}^2+2\theta_{pq}^2)
   \nonumber \\
  8\mathcal{F}_{\tau\tau^V\tau^V}^4&=&-\bar{\beta}(\bar{\mu}\alpha_p\gamma_q+\bar{\gamma}\gamma_p\beta_q)
   -2\bar{\beta}\theta_{pq}(\alpha_q\beta_q+\alpha_p\beta_p)
   -2\gamma_p\gamma_q(\alpha_p^2+\beta_p^2+\alpha_q^2+\beta_q^2+\frac{1}{2}\bar{\beta}^2+2\theta_{pq}^2) \nonumber \\ && \;
   -2(\alpha_p\alpha_q+\beta_p\beta_q)(\gamma_p^2+\gamma_q^2+2\theta_{pq}^2) \nonumber \\
  8\mathcal{F}_{\tau\tau^V\tau^V}^5&=&4\theta_{pq}\gamma_p\gamma_q(\alpha_p\alpha_q+\beta_p\beta_q)
   +2\bar{\beta}\gamma_p\gamma_q(\alpha_p\beta_p+\alpha_q\beta_q)+2\bar{\beta}\alpha_p\beta_q(\gamma_p^2+\gamma_q^2+2\theta_{pq}^2)
    \nonumber \\
  8\mathcal{F}_{\tau\tau^V\tau^V}^6&=&-4\bar{\beta}\theta_{pq}\alpha_p\gamma_p\beta_q\gamma_q. \nonumber
\eea

The cross-correlation with the anisotropic pressure is recovered with the operator $\mathcal{A}_{ijklmn}=-Q_{ij}(\mathbf{k})\hat{p}_{(k}P_{l)a}(\mathbf{p})\hat{q}_{(m}P_{n)a}(\mathbf{q})$; the eventual result is
\be
  \mathcal{F}_{\tau^S\tau^V\tau^V}=-\sum_{n=1}^7\mathcal{F}_{\tau^S\tau^V\tau^V}^{n}
\ee
\noindent with
\bea
    8\mathcal{F}_{\tau^S\tau^V\tau^V}^1&=&6\theta_{pq} \nonumber \\
    -8\mathcal{F}_{\tau^S\tau^V\tau^V}^2&=&4\gamma_p\gamma_q \nonumber \\
    -8\mathcal{F}_{\tau^S\tau^V\tau^V}^3&=&
      \left(\beta_p\gamma_q+\beta_q\gamma_p\right)\overline{\mu}
      +(\alpha_p\gamma_q+\alpha_q\gamma_p)\overline{\gamma}
      +\left(3\theta_{kp}\gamma_q+3\theta_{kq}\gamma_p\right)\gamma_k
      \nonumber \\ && \;
      +\theta_{pq}\left(6\theta_{kp}^2+6\theta_{kq}^2+4\theta_{pq}^2+3\alpha_k^2+3\beta_k^2+2\alpha_p^2
        +2\beta_p^2+2\gamma_p^2+2\alpha_q^2+2\beta_q^2+2\gamma_q^2+\beta_b^2\right) \nonumber \\
    8\mathcal{F}_{\tau^S\tau^V\tau^V}^4&=&
      4\theta_{pq}^2\left(3\theta_{kp}\theta_{kq}+\alpha_p\alpha_q+\beta_p\beta_q+\gamma_p\gamma_q\right)
      +\overline{\beta}\theta_{pq}\left(3\alpha_k\beta_k+2\left(\alpha_p\beta_p+\alpha_q\beta_q\right)\right)
        \nonumber \\ && \;
      +6\theta_{pq}\left(\theta_{kp}\left(\alpha_k\alpha_p+\beta_k\beta_p\right)
        +\theta_{kq}\left(\alpha_k\alpha_q+\beta_k\beta_q\right)\right)
        \nonumber \\ && \;
      +\gamma_p\gamma_q\left(3\alpha_k^2+3\beta_k^2+2\alpha_p^2+2\beta_p^2+2\alpha_q^2+2\beta_q
        +6\theta_{kp}^2+6\theta_{kq}^2+\overline{\beta}^2\right)
        \nonumber \\ && \;
      +2\left(\gamma_p^2+\gamma_q^2\right)\left(\alpha_p\alpha_q+\beta_p\beta_q+3\theta_{kp}\theta_{kq}\right)
      +\overline{\beta}\left(\alpha_p\gamma_q\overline{\mu}+\beta_q\gamma_p\overline{\gamma}\right)
      +3\left(\alpha_k\gamma_p\overline{\gamma}\theta_{kq}+\beta_k\gamma_q\overline{\mu}\theta_{kp}\right)
        \nonumber \\ && \;
      +3\gamma_k\left(\alpha_k\alpha_p\gamma_q+\beta_k\beta_q\gamma_p\right) \nonumber \\
    -8\mathcal{F}_{\tau^S\tau^V\tau^V}^5&=&
      4\theta_{pq}^2\left(\alpha_p\beta_q\overline{\beta}+3\theta_{kq}\alpha_k\alpha_p+3\theta_{kp}\beta_k\beta_q\right)
      +4\left(\theta_{pq}\left(\alpha_p\alpha_q+\beta_p\beta_q+3\theta_{kp}\theta_{kq}\right)\gamma_p\gamma
        \right. \nonumber \\ && \; \left.
        +3\alpha_k\beta_k\left(\alpha_p\beta_p+\alpha_q\beta_q\right)\right)
      +6\theta_{kp}\left(\left(\alpha_k\alpha_p+\beta_k\beta_p\right)\gamma_p\gamma_q
        +\beta_k\beta_q\left(\gamma_p^2+\gamma_q^2\right)\right)
        \nonumber \\ && \;
      +6\theta_{kq}\left(\left(\alpha_k\alpha_q+\beta_k\beta_q\right)\gamma_p\gamma_q
        +\alpha_k\alpha_p\left(\gamma_p^2+\gamma_q^2\right)\right)
      +\left(1\alpha_p\beta_p+1\alpha_q\beta_q+3\alpha_k\beta_k\right)\gamma_p\gamma_q\overline{\beta}
        \nonumber \\ && \;
      +2\alpha_p\beta_q\overline{\beta}\left(\gamma_p^2+\gamma_q^2\right)
      +3\alpha_k\beta_k\left(\alpha_p\gamma_q\overline{\mu}+\beta_q\gamma_p\overline{\gamma}\right)
      \nonumber \\
    8\mathcal{F}_{\tau^S\tau^V\tau^V}^6&=&
      12\theta_{pq}^2\alpha_k\alpha_p\beta_k\beta_q
      +4\theta_{pq}\gamma_p\gamma_q\left(\alpha_p\beta_q\overline{\beta}+3\theta_{kq}\alpha_k\alpha_p
       +3\theta_{kp}\beta_k\beta_q\right)
        \nonumber \\ && \;
      +6\alpha_k\beta_k\gamma_p\gamma_q\left(\alpha_p\beta_p+\alpha_q\beta_q\right)
      +6\alpha_k\beta_k\alpha_p\beta_q\left(\gamma_p^2+\gamma_q^2\right) \nonumber \\
    -8\mathcal{F}_{\tau^S\tau^V\tau^V}^7&=&12\theta_{pq}\alpha_k\alpha_p\beta_k\beta_q\gamma_p\gamma_q. \nonumber
\eea

The full tensor correlation, $\langle\tau^T_{ab}\tau^T_{bc}\tau^T_{ac}\rangle$, has not been calculated, but it can be found by the application of $\mathcal{A}_{ijklmn}=\left(P_{a(i}(\mathbf{k})P_{j)b}(\mathbf{k})-\frac{1}{2}P_{ij}(\mathbf{k})P_{ab}(\mathbf{k})\right)\left(P_{b(k}(\mathbf{p})P_{l)c}(\mathbf{p})-\frac{1}{2}P_{kl}(\mathbf{p})P_{bc}(\mathbf{p})\right)\left(P_{a(m}(\mathbf{q})P_{n)c}(\mathbf{q})-\frac{1}{2}P_{mn}(\mathbf{q})P_{ac}(\mathbf{q})\right)$. If we consider first the colinear case, this reduces immediately to a product of projection tensors on $\mathbf{k}$. Employing the symmetries of $\mathcal{A}_{ijklmn}$ in $\{ij\}$, $\{kl\}$ and $\{mn\}$ and of $\mathcal{B}_{ijklmn}$ in $\{ik\}$, $\{jm\}$ and $\{ln\}$ one may then demonstrate that this vanishes identically. The general case is not straight forward and we defer it to a later study.

\subsection{Flat spectrum results}
\begin{figure}
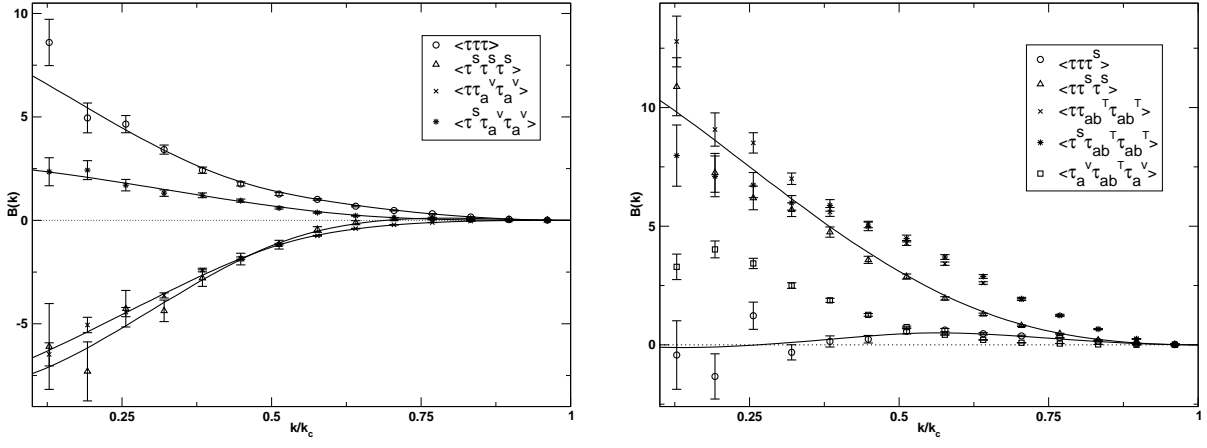
\begin{center}
\includegraphics{3Points_n0_1.eps}\qquad\includegraphics{3Points_n0_2.eps}
\caption{Here we show the various colinear bispectra for the flat spectrum $n=0$ generated from a collection of realizations.  Where the analytic results have been calculated, they agree well with the numerical results.}
\label{3Point_n0}\end{center}\end{figure}

As was the case for the power spectra, there are two very different spectral regimes for the bispectra.  For flat spectra, $n>-1$, the integrals are dominated by the highest $k$ modes around the cutoff scale.  For these spectral indices, the bispectra become independent of $k$ when
 $k\ll k_c$ and the analytic expressions are straight forward to integrate.  Indeed, we can do them exactly in the limit $k\ll k_c$; we present these results below both for a generic geometry and for the colinear case. For convenience we define the generic geometry using the two angle cosines $\alpha_k=\cos(\bar{\theta})$ and $\theta_{kq}=-\cos(\xi_{kq})$ from which our specification of $r$ and $\phi$ may be recovered -- see Figure \ref{BispectraGeometry}.

In the general case, setting $B=A^2k_c^{3(n+1)}/3(n+1)$ with $A$ the amplitude of the magnetic field power spectrum, we find
\bea
  \langle\tau(\mathbf{k})\tau(\mathbf{p})\tau(\mathbf{q})\rangle&=&B\pi\delta(\mathbf{k}+\mathbf{p}+\mathbf{q})
    \nonumber \\
  \langle\tau(\mathbf{k})\tau(\mathbf{p})\tau^S(\mathbf{q})\rangle&=&0
    \nonumber \\
  \langle\tau(\mathbf{k})\tau^S(\mathbf{p})\tau^S(\mathbf{q})\rangle&=&
   \frac{21}{30}B\pi\left(2+6\cos(\phi)\cos(\xi_{kq})-3\cos^2(\xi_{kq})-3\cos^2(\phi)+6\cos^2(\phi)\cos^2(\xi_{kq})
    \right. \nonumber \\ && \; \left.
   -6\cos(\phi)\cos^3(\xi_{kq})-6\cos^3(\phi)\cos(\xi_{kq})
   +6\cos^3(\phi)\cos^3(\xi_{kq})\right)\delta(\mathbf{k}+\mathbf{p}+\mathbf{q})
    \nonumber \\
  \langle\tau^S(\mathbf{k})\tau^S(\mathbf{p})\tau^S(\mathbf{q})\rangle&=&
   \frac{17}{35}B\pi\left(1-3\cos^2(\phi)\cos^2(\xi_{kq})
   -3\cos(\phi)\cos(\xi_{kq})\sin^2(\phi)\sin^2(\xi_{kq})\right)\delta(\mathbf{k}+\mathbf{p}+\mathbf{q})
    \\
  \langle\tau(\mathbf{k})\tau_a^V(\mathbf{p})\tau_a^V(\mathbf{q})\rangle&=&
   -\frac{14}{15}B\pi\left(3\cos(\phi)\cos(\xi_{kq})+\sin^2(\phi)\sin^2(\xi_{kq})
   -3\cos^3(\phi)\cos(\xi_{kq})-3\cos(\phi)\cos^3(\xi_{kq})
    \right. \nonumber \\ && \;
   +4\cos^3(\phi)\cos^3(\xi_{kq})
   -\sin^2(\phi)\cos^2(\phi)\sin^2(\xi_{kq})-\sin^2(\phi)\sin^2(\xi_{kq})\cos^2(\xi_{kq})
    \nonumber \\ && \; \left.
   +4\cos^2(\phi)\cos^2(\xi_{kq})\sin^2(\phi)\sin^2(\xi_{kq})\right)\delta(\mathbf{k}+\mathbf{p}+\mathbf{q})
     \nonumber \\
  \langle\tau^S(\mathbf{k})\tau_a^V(\mathbf{p})\tau_a^V(\mathbf{q})\rangle&=&
   \frac{17}{105}B\pi\left(6\cos(\phi)\cos(\xi_{kq})
   -6\cos^3(\phi)\cos(\xi_{kq})-6\cos(\phi)\cos^3(\xi_{kq})-\sin^2(\phi)\sin^2(\xi_{kq})
    \right. \nonumber \\ && \;
   +8\cos^3(\phi)\cos^3(\xi_{kq})-2\sin^2(\phi)\sin^2(\xi_{kq})\cos^2(\xi_{kq})-2\sin^2(\phi)\sin^2(\xi_{kq})\cos^2(\phi)
    \nonumber \\ && \; \left.
   +8\cos^2(\phi)\cos^2(\xi_{kq})\sin^2(\phi)\sin^2(\xi_{kq})\right)\delta(\mathbf{k}+\mathbf{p}+\mathbf{q})
    \nonumber
\eea

Specialising these to the colinear case wherein $\phi=\xi_{kq}=0$ these reduce to
\bea
  \langle\tau(\mathbf{k})\tau(\mathbf{p})\tau(\mathbf{q})\rangle&=&B\pi\delta(\mathbf{k}+\mathbf{p}+\mathbf{q})
    \nonumber \\
  \langle\tau(\mathbf{k})\tau(\mathbf{p})\tau^S(\mathbf{q})\rangle&=&0
    \nonumber \\
  \langle\tau(\mathbf{k})\tau^S(\mathbf{p})\tau^S(\mathbf{q})\rangle&=&
    \frac{7}{5}B\pi\delta(\mathbf{k}+\mathbf{p}+\mathbf{q})
    \nonumber \\
  \langle\tau^S(\mathbf{k})\tau^S(\mathbf{p})\tau^S(\mathbf{q})\rangle&=&
   -\frac{34}{35}B\pi\delta(\mathbf{k}+\mathbf{p}+\mathbf{q})
   \\
  \langle\tau(\mathbf{k})\tau_a^V(\mathbf{p})\tau_a^V(\mathbf{q})\rangle&=&
   -\frac{14}{15}B\pi\delta(\mathbf{k}+\mathbf{p}+\mathbf{q})
     \nonumber \\
  \langle\tau^S(\mathbf{k})\tau_a^V(\mathbf{p})\tau_a^V(\mathbf{q})\rangle&=&
   \frac{34}{105}B\pi\delta(\mathbf{k}+\mathbf{p}+\mathbf{q})
    \nonumber
\eea

The bispectra that are derived from the simulated fields are heavily compromised by the grid-size; unlike the two-point case the three-point moments use only a restricted number of the modes, selected by the geometry chosen for the wavevectors. The result from a single realization is in most cases noise-dominated. To overcome this difficulty we have chosen to simulate a large number of different realizations, taking the mean signal and using their variance to provide an estimate for the errors involved. The results, for a flat power spectrum, a grid-size of $l_{\mathrm{dim}}=192$ (and a damping scale of $k_c=l_{\mathrm{dim}}/4.1$) and $1,500$ combined realizations, are plotted rebinned in Figures \ref{3Point_n0} with the numerically-integrated predictions overlaid. For simplicity we have concentrated on the colinear case, wherein $\mathbf{p}=\mathbf{k}$ and so $\mathbf{q}=-\left(\mathbf{k}+\mathbf{p}\right)$. We plot $k/k_c$ against $B(k)$ where $B(k)$ represents the colinear bispectra.

\subsection{Steep spectrum results}
\begin{figure}
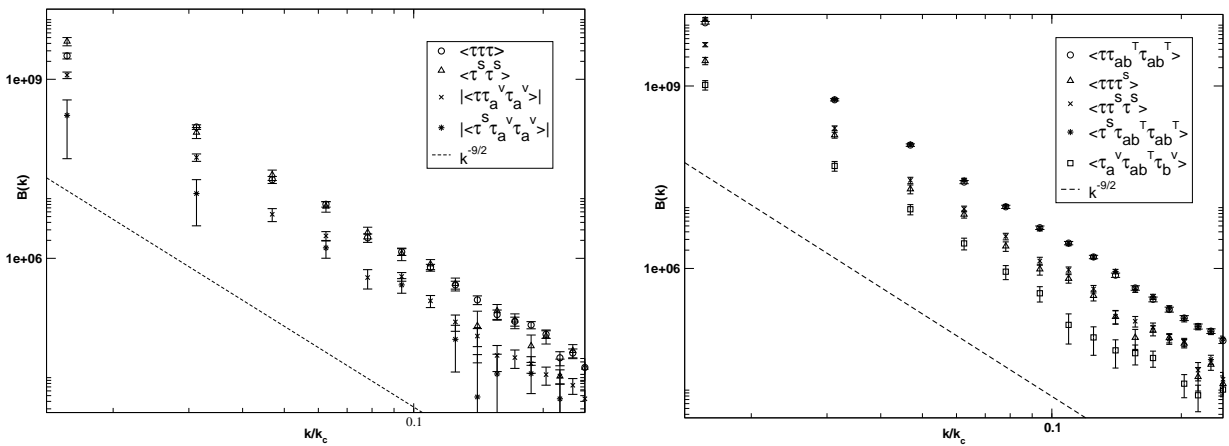
\begin{center}
\includegraphics{3Points_n-2.5_1.eps}\qquad\includegraphics{3Points_n-2.5_2.eps}
\caption{Here we show the colinear bispectra for a steep spectrum ($n=-2.5$.) The scaling is as expected naively, $\propto k^{3n+3} = k^{-9/2}$ (dotted line).}
\label{3Point_n-2.5}\end{center}\end{figure}

In the regime $n<-1$ matters are, as with the two-point case, complicated by the presence of numerous poles; the integrals are dominated by the volume lying between the poles.
Rather than attempt a solution, we use our realizations to calculate the bispectra.
We present the results for $n=-2.5$ in Figures \ref{3Point_n-2.5}.  There is a dependence on the gridsize for this spectrum due to the paucity of modes of appreciable power given the strongly red spectrum. We tested this by running three realizations with differing grid-sizes, keeping the total number of modes constant in each case constant; specifically we took $1,500$ simulations at $l_{\mathrm{dim}}=192$, $5000$ simulations at $l_{dim}=128$, and $40,000$ simulations at $l_{dim}=64$. As might be expected we found a suppression for the case wherein $l_{dim}=64$ -- due to scarcity of modes at low $k$ -- but there was good agreement between the other two cases. We have plotted the results from the $l_{dim}=192$ run for greatest dynamic range, again rebinning into $64$ bins.

With this highly-tilted spectrum we see, as with the two-point case, that the features of the magnetic spectrum at the cut-off scale apparent in the flat case are washed out by the spectral tilt. The predominant shape again is the $k$-dependence we expect from the radial component of the integral; in this case $B(k)\propto k^{3(n+1)}$. (These are plotted for comparison.) Again, as with the two-point case there is an infrared suppression. The magnitudes of the bispectra fall into three close bands; the strongest are the correlations between the scalars and the tensors, while the middle-band is composed of the scalar auto- and cross-correlations. The weakest bispectra are those involving correlations with the vectors and, in some cases, the $1-\sigma$ error bars are greater than the mean value; such points have been removed from these plots for aesthetic purposes. The $\langle\tau^V_a\tau^T_{ab}\tau^V_b\rangle$ correlation is particularly weak.

\section{Discussion and Conclusions}
 We have studied, analytically and via realizations, some of the higher point correlations for tangled large-scale magnetic fields.  The analysis is obviously highly model-dependent, and extending the analytic results to other models could be very difficult.  However, it would be a simple matter to generalize the numerical realizations to perform the same analysis for a wide variety of time-independent models; all we require is the statistical distribution of the underlying field, the power-spectrum and the form of the stress-energy tensor. For example, while we have assumed the simplest -- and perhaps most instructive -- case of a magnetic field with Gaussian statistics, it would be a simple matter to employ a $\chi^2$ probability distribution for a magnetic field, which is physically motivated by the creation mechanism considered by Matarrese \emph{et. al.} \cite{MatarreseEtAl04}. (Since our code is time-independent the interpretation would be of the final field immediately prior to recombination.)  The realizations will however be limited by the narrow dynamic range allowed in the computation.

In our particular case of a tangled primordial magnetic field with a power-law spectrum, we have demonstrated that we can recover the 1-, 2- and 3-point statistics from simulations and with an excellent agreement to our analytic predictions. At the one-point level we have not only verified that the magnetic energy density follows a $\chi^2$ probability distribution function (as expected given that it is directly the square of the underlying Gaussian magnetic field), but that the anisotropic pressure is also non-Gaussian with a significantly more complex relation to the fields. There is also a spectral dependence on its probability distribution function affecting the skewness, which swaps sign as one passes through $n=-3/2$. The kurtosis, while exhibiting a spectral dependence, remains positive.

At the two-point level we have calculated the scalar, vector and tensor auto-correlations, as well as the scalar-cross correlation. For the scale-invariant spectrum we confirm the power-spectra with the expected ratios for scales longer than the cutoff; we also see that the scalar cross-correlation vanishes on large scales but is not in general entirely zero for modes close to the cutoff scale. For a highly-tilted spectrum, the cutoff scale is less important and the spectra behave with power law behavior and with the relative ratios approximately constant. The $k$-dependence is the power-law that would be expected from a naive point of view. The surprise is that in this regime the scalar cross-correlation no longer vanishes on large-scales; rather, the correlation remains roughly constant at $\langle\tau\tau^S\rangle/\sqrt{\langle\tau\tau\rangle\langle\tau^S\tau^S\rangle}\approx 0.7.$ Thus, while the isotropic and anisotropic pressures are indeed correlated, they are not perfectly correlated.

At the three-point level we considered a number of rotationally-invariant bispectra, concentrating for simplicity on the colinear case. We find significant non-Gaussianities -- in excellent agreement with the analytic predictions.  For the flat spectrum, these approach fixed ratios on large scales.  These can be both positive or negative, or even zero.  As in the two-point case, some qualitative aspects change when we consider a strongly-tilted magnetic power spectrum; features arising from the cutoff scale tend to disappear, leaving instead a simple power-law drop off.  The relative ratios of the bispectra also change, and even their relative signs differ.  The correlation between the isotropic pressure squared and the anisotropic pressure, which vanishes for scale invariant spectra, becomes non-zero.

It remains for future work \cite{BrownCrittenden2} to fold these results in with the transfer functions for magnetized cosmologies and calculate the non-Gaussianities expected to be imprinted upon the cosmic microwave sky. We can then consider how such signals might be used to constrain the properties of a magnetic field of this type. It would also obviously be straight-forward to consider different magnetic power spectra and statistics.  While it is too early to speculate what these will discover, the higher order correlations in the sources will certainly lead to similar higher order correlations in the CMB observables, including perhaps cross correlations between the polarization modes such as $\langle E^2 B^2 \rangle$. 

The techniques used in this paper could be applied to a broad variety of models.
Moreover, we have here to our knowledge presented the first calculations of cross-correlations between, for example, scalars and tensors, and demonstrated that they can be of an equal magnitude to scalar auto-correlations.  This has great potential relevance to the wider field of sources with non-linear stress-energy tensors, or sources with non-Gaussian initial conditions. For example, it would be interesting to compare our results to the same moments evaluated for defect models, particularly given the current resurgence of interest into networks of cosmic strings.

\begin{acknowledgments}
We wish to thank A. Lewis, R. Maartens, K. Subramanian and K. Dimopoulos for useful discussions.
\end{acknowledgments}

\bibliography{paper}

\end{document}